\documentclass[aps,pra,twocolumn,superscriptaddress,showpacs]{revtex4}

\usepackage{graphicx}

\begin{document}

\title{A Mott Insulator of Ultracold Alkaline-Earth-Like Atoms}

\author{Takeshi Fukuhara}
\affiliation{Department of Physics, Graduate School of Science, Kyoto University, Kyoto 606-8502, Japan}

\author{Seiji Sugawa}
\affiliation{Department of Physics, Graduate School of Science, Kyoto University, Kyoto 606-8502, Japan}

\author{Masahito Sugimoto}
\affiliation{Department of Physics, Graduate School of Science, Kyoto University, Kyoto 606-8502, Japan}

\author{Shintaro Taie}
\affiliation{Department of Physics, Graduate School of Science, Kyoto University, Kyoto 606-8502, Japan}

\author{Yoshiro Takahashi}
\affiliation{Department of Physics, Graduate School of Science, Kyoto University, Kyoto 606-8502, Japan}
\affiliation{CREST, Japan Science and Technology Agency, Kawaguchi, Saitama 332-0012, Japan}

\date{\today}

\begin{abstract}
The transition from a superfluid to a Mott insulator (MI) phase has been observed in a Bose-Einstein condensate (BEC) of ytterbium (Yb) atoms in an optical lattice. An all-optically produced BEC of $^{174}$Yb atoms was loaded into three-dimensional optical lattices produced by a 532 nm laser beam. The interference pattern was measured after releasing the quantum gas from the trapping potential. As increasing the optical lattice depth, we observed the disappearance of the interference patterns, which is a signature of entering the MI regime. This result is an important step into studies by using a combination of the MI state and the ultranarrow optical transition of ultracold alkaline-earth-like atoms. 
\end{abstract}


\maketitle
A atomic quantum gas in a periodic potential is an important and promising system for a variety of studies. Since the first observation of the phase transition from a superfluid to a Mott insulator (MI) in a atomic Bose-Einstein condensate (BEC) of Rb atoms held in a three-dimensional (3D) optical lattice \cite{Greiner02}, strongly correlated quantum systems using ultracold atoms in optical lattices have been extensively studied. In addition, the MI phase with one atom per site is an important starting point for implementing quantum information processing based on ultracold atoms \cite{Jaksch99}. So far, however, the MI phase has been realized only in alkali atoms \cite{Greiner02,Xu05}. Here we report the observation of the transition from a superfluid to a MI phase in ytterbium (Yb), an alkaline-earth-like atom.

It is important to realize a MI phase in Yb for the following reasons. First, Yb is an excellent candidate for an optical lattice clock \cite{Takamoto05,Taichenachev06}. The absolute frequency of the $^1S_0$$-$$^3P_0$ transition in bosonic $^{174}$Yb atoms has been recently determined within a fractional uncertainty of $1.5 \times 10^{-15}$ \cite{Poli08}. The main contribution to the uncertainty is the density shift due to cold collisions between ultracold $^{174}$Yb atoms in a 1D optical lattice. A method to eliminate the collisional frequency shifts is to load bosonic atoms into 3D optical lattices \cite{Akatsuka08}, and the preparation of the MI phase with single atom per site leads to the largest signal-to-noise ratio. Second advantage arises from the possibility of high-resolution spectroscopy using the ultranarrow transitions of $^1S_0$$-$$^3P_0$ and $^1S_0$$-$$^3P_2$. From high-resolution laser spectroscopy of a BEC in $^{174}$Yb, density-dependent collisional shifts have been found to be large \cite{Uetake,Yamaguchi08}, and thus we can spectroscopically investigate a shell structure of the MI phase and probe the interaction energy by using the ultranarrow transitions with increased resolution. Third, two fermionic isotopes ($^{171}$Yb with nuclear spin $I=1/2$ and $^{173}$Yb with $I=5/2$) have been cooled down to below the Fermi temperature \cite{Fukuhara08,Fukuhara07a}, and intriguing systems of ultracold fermionic gases in optical lattices can be provided by using basically the same procedure described here. They are especially important because interesting schemes for quantum computing based on the nuclear spin state in alkaline-earth-like atoms are recently proposed \cite{Shibata,Daley08}. 

\begin{figure}[b]
\begin{center}
\includegraphics[width=\linewidth]{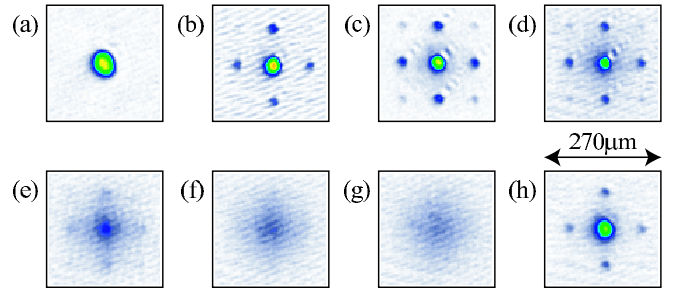}
\end{center}
\caption{(color online). Absorption images of matter-wave interference patterns after releasing atoms from an optical lattice potential with a potential depth of (a) 0$E_R$, (b) 5$E_R$, (c) 8$E_R$, (d) 11$E_R$, (e) 14$E_R$, (f) 17$E_R$, and (g) 20$E_R$ ($E_R = \hbar ^2 k_L^2 / 2 m$ is the recoil energy with $k_L = 2\pi / \lambda _ L$ and the atomic mass $m$). The decrease in the visibility of the interference pattern is observed as the lattice depth increases. (h) Interference pattern is restored after preparing a MI phase at a lattice depth of 20$E_R$ and decreasing the potential to 5$E_R$ within 15 ms. The data are averaged over three measurements. \label{transition}}
\end{figure}
The transition from a superfluid to a MI phase in a BEC of alkaline-earth-like $^{174}$Yb atoms is observed in optical lattices created by lasers at a wavelength of $\lambda _L = 532$ nm. A multiple matter-wave interference pattern is measured after suddenly releasing the atoms from the optical lattice potential and a subsequent time-of-flight period of 10ms (Fig. \ref{transition}). As increasing the optical lattice depth $V_0$, we observe the disappearance of the interference patterns, which is a signature of entering the MI regime. We also observe a restoration of the interference pattern after ramping up to a lattice depth of 20$E_R$, where the system is in a MI phase, and ramping down the potential to 5$E_R$, where the system is in a superfluid phase. 

The experiment is carried out in our new apparatus with 20 viewports providing excellent optical access. The background gas pressure in the chamber is $\sim 3 \times 10^{-11}$ Torr. The experimental procedure for producing a $^{174}$Yb BEC is basically the same as in our previous work \cite{Takasu03,Fukuhara07b}. Yb atoms are decelerated by a Zeeman slower with the strong transition ($^1S_0$$-$$^1P_1$, a wavelength of 399 nm and a linewidth of 29 MHz) and then loaded into a magneto-optical trap (MOT) with the intercombination transition ($^1S_0$$-$$^3P_1$, a wavelength of 556 nm and a linewidth of 182 kHz). The MOT beams are created by frequency doubling of a 1112-nm fiber laser \cite{Uetake08}. The laser-cooled Yb atoms are transferred from the MOT to an crossed far-off-resonance trap (FORT). 
\begin{figure}
\begin{center}
\includegraphics[width=\linewidth]{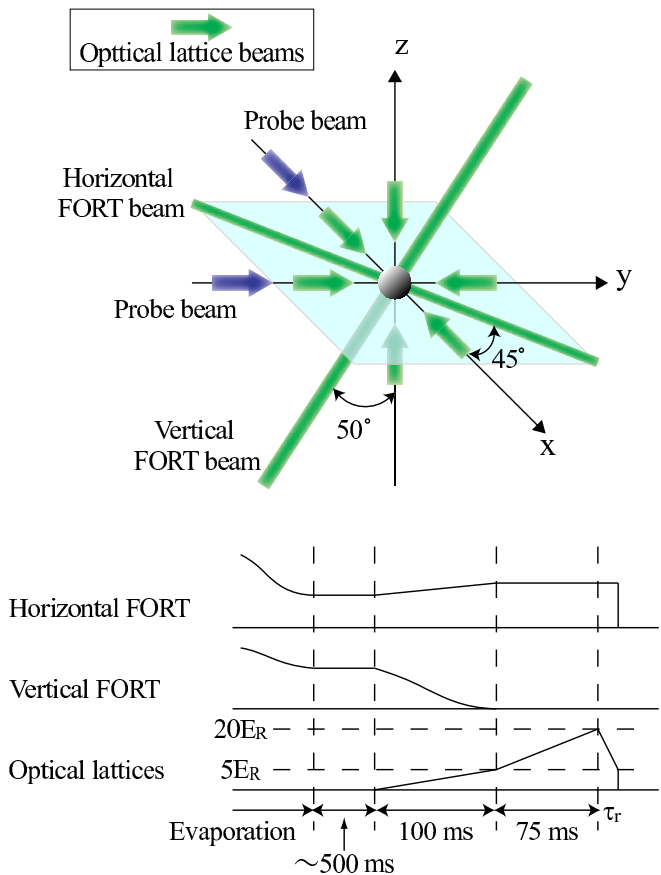}
\end{center}
\caption{(color online). Schematic diagram of experimental setup (upper) and experimental procedure to load the atoms into optical lattices (lower). Absorption images can be taken along the two orthogonal lattice axes to confirm 3D interference pattern. \label{lattice}}
\end{figure}
In our previous experiments, the crossed FORT is generated by two laser beams, which are oriented horizontally and vertically \cite{Takasu03,Fukuhara07a,Fukuhara07b}. In this experiment, however, the vertical FORT beam, perpendicular to the horizontal FORT beam, is at an angle of approximately 50$^{\circ }$ from the vertical z axis (Fig. \ref{lattice}). It is noted that the configuration of the beam is not so important since the main role of the second FORT beam is to increase the atom density by loading atoms into the crossing region. Evaporative cooling is carried out by lowering the potential depth of the horizontal FORT beam while the depth of the vertical FORT beam kept constant. At the final stage of evaporation, where the temperature of the cloud is near the BEC transition temperature, the power of the vertical FORT beam is also decreased to suppress the three-body recombination atom loss by lowering the atom density. After evaporation the atoms are held in the trap for typically 500 ms, resulting in a quasipure BEC with no discernible thermal cloud. 

A $^{174}$Yb BEC with up to $3 \times 10^4$ atoms is loaded into 3D optical lattices. The lattice potential is formed by three orthogonal, retroreflected laser beams (Fig. \ref{lattice}). The beams are produced by the same laser (10 W diode-pumped solid-state lasers operating at 532 nm, Coherent Verdi-V10) used for the horizontal FORT beam; the first-order diffraction beam of an acousto-optic modulator (AOM) is used for the horizontal FORT beam and the transmitted beam for the optical lattices. The linewidth of the laser is less than 5 MHz. In order to eliminate interferences between different beams, the three lattice beams and the horizontal FORT beam are frequency shifted at least 5 MHz relative to each other by using AOMs and the polarization of the three lattice beams is orthogonal to each other.  The three beams are spatially filtered with a single-mode optical fiber before being focused to a $1/e^2$ beam waist of $\sim$50 $\mu$m at the position of the BEC, and the intensities are stabilized by using AOMs. The lattice depth is calibrated by observing the diffraction caused by a pulsed optical lattice \cite{Ovchinnikov99}.  

The lattice depth is linearly increased to 5$E_R$ in 100 ms, during which the intensity of the vertical FORT beam is decreased to zero. Here, the intensity of the horizontal FORT beam is also slightly increased to keep constant the trap confinement along the gravity. At this point, there is an additional external harmonic potential by the horizontal FORT beam and the lattice beams with a gaussian shape. The radial trapping frequencies of the horizontal FORT beam is measured to be $2\pi \times$160 Hz (the horizontal direction) and $2\pi \times$240 Hz (the vertical direction) by exciting the center-of-mass motion. To confirm that the condensate is loaded without decoherence, we observe the interference pattern in the density distribution of the atoms suddenly released from the optical lattices and the FORT (Fig. \ref{transition}(b)). Subsequently, the optical lattice is linearly ramped up to maximum potential of 20$E_R$ in 75 ms. As the lattice depth is increased, the interference pattern disappears (Fig. \ref{transition}(b)$-$(g)), which indicates the transition to the MI state, accompanied by the loss of phase coherence \cite{Greiner02}. After forming the MI phase at the lattice depth of 20$E_R$ and ramping down the potential depth to 5$E_R$, the revival of the interference pattern is also observed. We measure the width of the center peak for various ramp-down times $\tau_r$ and obtain the time scale of 0.4 ms for restoration of coherence. The time is comparable to the tunneling time scale $\hbar/J \sim$ 0.6 ms at a potential depth of 5$E_R$ \cite{Greiner02}, where $J$ is the tunneling matrix element.

\begin{figure}
\begin{center}
\includegraphics[width=\linewidth]{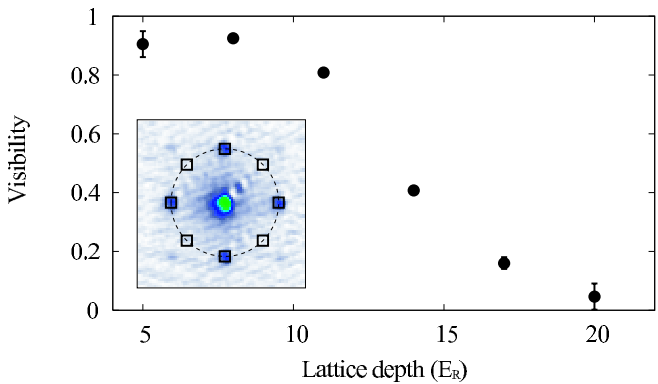}
\end{center}
\caption{(color online). Visibility as a function of lattice depth. Each data point is the average of three measurements. The error bars represent only the standard errors of the measurements. The inset shows the regions used for evaluating the visibility (see text). \label{visibility}}
\end{figure}

To quantitatively evaluate the interference fringes, we define the visibility of the interference pattern as $\mathcal{V}=\left( N_{max} - N_{min} \right) / \left( N_{max} + N_{min} \right)$ \cite{Gerbier05}. Here $N_{max}$ is the sum of the number of atoms in the regions of the first order interference maxima and $N_{min}$ is the sum of the number of atoms in equivalent regions at the same distance from the center peak along the diagonals (see the inset of Fig. \ref{visibility}). We measure the visibility $\mathcal{V}$ as a function of the lattice depth $V_0$ (Fig. \ref{visibility}). At the point where the condensate are loaded into the optical potential with the lattice depth $V_0=$5$E_R$, the visibility is $\sim 1$. We can see the decrease in the visibility at a lattice depth higher than 11$E_R$. The visibility becomes almost zero at the maximum potential of 20$E_R$. 

The ultracold Bose atoms in the optical lattices can be described by the Bose-Hubbard Hamiltonian \cite{Fisher89,Jaksch98}. According to the mean-field theory for the homogenous systems, the critical value for the phase transition to the MI state with $n$ atoms per lattice site is estimated to be \cite{Krauth92}
\begin{equation}
\frac{U}{J} = z \left\{2 n + 1 + 2 \sqrt{n \left( n+1 \right)} \right\},
\end{equation}
where $U$ is the on-site interaction energy and $z$ is the number of nearest neighbors ($z=6$ for simple cubic 3D lattices). We use the approximate expressions $U/E_R = 5.97 \left(a_s / \lambda _L \right) \left(V_0 / E_R \right)^{0.88}$, and $J/E_R = 1.43 \left(V_0 / E_R \right)^{0.98} \exp\left(-2.07\sqrt{V_0 / E_R} \right)$ \cite{Gerbier05}, with the scattering length $a_s = 5.55$ nm \cite{Kitagawa08}. By using these formulas, we find that the critical values are $V_0=$11$E_R$, 13$E_R$, 14$E_R$ and 15$E_R$ for $n=$1, 2, 3, and 4, respectively. These values are in good agreement with the reduction in the measured visibility above 11$E_R$. 

The peak occupation number can be simply estimated assuming zero temperature and no tunneling \cite{DeMarco05}. We define the parameter
\begin{equation}
\xi =\frac{3 N d^3}{4 \pi }\left(\frac{m \bar{\omega }^2}{2U} \right)^{3/2},
\end{equation}
where $N$ is the total number of atoms, $d$ is the lattice spacing ($d=\lambda _L / 2$), and $\bar{\omega}=(\omega_1 \omega_2 \omega_3)^{1/3}$ is the geometrical average of the trap frequencies of the external harmonic confinement. If the condition 
\begin{equation}
\Sigma _{k=0}^{n_0 - 1} k^{3/2} < \xi < \Sigma _{k=0}^{n_0} k^{3/2}
\end{equation}
is satisfied, then the peak occupation number is $n_0$. In our case of $3 \times 10^4$ atoms in the lattice potential with $V_0=$20$E_R$, where the trapping frequencies are estimated to be $2\pi \times$(120, 200, 270) Hz, the parameter $\xi$ is calculated to be 3.3, and thus the peak occupation number should be $n_0=2$.

In order to experimentally investigate the occupation numbers, spectroscopic studies are proven to be powerful methods \cite{Campbell06}. Here we discuss the possibility of the spectroscopic study of the MI shells of Yb atoms. While the spectroscopy of the MI shells by using two-photon rf spectroscopy for Rb atoms is recently demonstrated \cite{Campbell06},  optical spectroscopy with increased resolution can be conducted for $^{174}$Yb atoms by using the ultranarrow optical transition of $^1S_0$$-$$^3P_2$. In order to measure the structure by using density-dependent collisional shifts of the transition from the state $|1>$ to the state $|2>$, the difference between $a_{11}$ and $a_{12}$ is important. Here $a_{11}$ is the scattering length between two atoms in the $|1>$ state and $a_{12}$ is the scattering length between two atoms in states $|1>$ and $|1>$. Concerning the MI shells, the collisional frequency shifts for the $n$ and $n-1$ MI phases are different by $\delta\nu =U/h\left[\left(a_{12} - a_{11} \right) / a_{11} \right]$ \cite{Campbell06}. For the transition of $^1S_0$$-$$^3P_2$ in $^{174}$Yb, the scattering lengths are $a_{11}=5.55$ nm and $a_{12} = - 33$ nm \cite{Yamaguchi08}. At a lattice potential with 20$E_R$, the difference in the collisional frequency shift $\delta\nu$ is estimated to be 24 kHz, which can be well resolved by using the ultranarrow transition of about 10 mHz linewidth. For investigation of the site occupation, this spectroscopic technique is quite useful. 

In conclusion, we have observed the phase transition from the superfluid to the MI state in a $^{174}$Yb BEC held in 3D optical lattices. As the lattice depth is increased, the disappearance of the interference pattern is observed above a potential depth of 11$E_R$, which is consistent with the prediction based on the mean-field theory. 

We acknowledge S. Uetake for experimental assistance. This work was partially supported by Grant-in-Aid for Scientific Research of JSPS (18204035) and the Global COE Program "The Next Generation of Physics, Spun from Universality and Emergence" from the Ministry of Education, Culture, Sports, Science and Technology (MEXT) of Japan. S. S. acknowledges support from JSPS.

\end{document}